\documentclass[preprint,showpacs,preprintnumbers,amsmath,amssymb,prl]{revtex4}
\usepackage[dvips]{graphicx}
\usepackage{dcolumn}         
\usepackage{bm}              

\begin{document}

\title{Structure and phase boundaries of compressed liquid hydrogen}

\author{Isaac Tamblyn} 
\email{itamblyn@dal.ca}
\author{Stanimir A. Bonev} 
\email{stanimir.bonev@dal.ca}
\affiliation{Department of Physics, Dalhousie University, Halifax, NS, B3H 3J5, Canada}
              
\date{\today}

\begin{abstract}

We have mapped the molecular-atomic transition in liquid hydrogen using first principles molecular dynamics. We predict that a molecular phase with short-range orientational order exists at pressures above 100 GPa. The presence of this ordering and the structure emerging near the dissociation transition provide an explanation for the sharpness of the molecular-atomic crossover and the concurrent pressure drop at high pressures. Our findings have non-trivial implications for simulations of hydrogen; previous equation of state data for the molecular liquid may require revision. Arguments for the possibility of a $1^{st}$ order liquid-liquid transition are discussed.

\end{abstract}


\maketitle

Constructing an accurate picture of high pressure hydrogen continues to present a challenge for both experiment and theory. In recent years, many studies have focused on the behavior of hydrogen at elevated temperatures. Topics of interest include the turnover of the melting line \cite{bonev_nature_2004,shanti_prl_2008,eremets_jetp_2009}, the molecular-atomic transition \cite{scandolo_pnas_2003,bonev_prb_2004,delaney_prl_2006, vorberger_prb_2007, caspersen}, a semiconductor to metallic transition \cite{weir_prl_1996, nellis_prb_1999}, as well as a host of temperature-induced dissociation and ionization phenomena. Despite considerable attention, many important questions about the chemical and physical properties of hydrogen remain unanswered. A complete description of the molecular-atomic transition has not been established. The relation between metallization and dissociation is still not fully understood. Effects due to impurities (e.g. helium) \cite{vorberger_prb_2007} and the accuracy of equation of state (EOS) data are particularly important for determining the structure of gas giants such as Jupiter\cite{cassen_springer_2006}.

This Letter is focused on a topic which has hitherto received little attention - the structure of the dense hydrogen liquid above the melting curve and near the dissociation transition. Based on first principles molecular dynamics (FPMD) simulations, we predict a liquid phase that exhibits complex short-range orientational order. This finding is significant, as the properties of this phase provide a physical explanation for some of the phenomena mentioned above.

\begin{figure}[tbh]
  \hspace*{-5mm}
  \includegraphics[width=0.5\textwidth, clip]{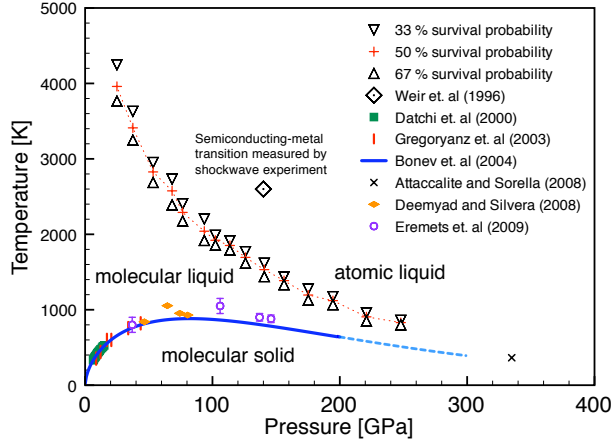} 
  \caption{\label{phase}Phase diagram of dense hydrogen. Triangles are upper and lower boundaries of the computed molecular-atomic transition. The $T$ range over which the transition takes place decreases with pressure, consistent with previous observations \cite{bonev_prb_2004,vorberger_prb_2007,caspersen,miguel}. Experimental \cite{datchi_prb_2000, gregoryanz_prl_2003, shanti_prl_2008, eremets_jetp_2009} and theoretical \cite{bonev_nature_2004} data (extrapolation for $P > 200$ GPa) of the melting line are shown along with the semiconducting-metallic transition \cite{weir_prl_1996, nellis_prb_1999}. Uncertainties in the $T$ of this measurement were estimated to be 30\% \cite{nellis_prb_1999}. QMC simulations \cite{attaccalite_prl_2008} confirm the stability of the liquid near 320 GPa and 364 K. }
\end{figure}

We have performed FPMD simulations using DFT-GGA \cite{kohn_sham,PBE}, mapping the molecular-atomic transition over a large pressure ($P$) and temperature ($T$) range (Fig.~\ref{phase}). Simulations were carried out using the Born Oppenheimer implementation in the CPMD code \cite{cpmd}. In this method, the electronic density is optimized to its ground state at each ionic step. We used a local Troullier-Martins pseudopotential with a 100 Ry planewave cutoff and $\Gamma$-point sampling of the Brillouin zone. Molecular dynamics (MD) simulations were carried out in the $NVT$ ensemble, where a Nos\'e-Hoover thermostat was used to control the ionic temperature. The supercell in our simulations consisted of 256 atoms. We performed checks on larger cells of 512, 768, and 1024 atoms; discussion of these results follow. An MD time-step of 16 a.u. (0.387 fs) was used for all simulations, with checks performed using a time-step of 8~a.u. Observables such as the pair correlation function, stress tensor, self-diffusion coefficient, and spacial distribution functions (discussed below), were all found to be well converged. In all cases, simulations were allowed to run for at least 2 ps and some as long as 15 ps. For the densities considered here, the system rapidly reaches equilibrium and these simulation times are sufficient. Calculations of the polarizability tensor were produced by extracting a large number of independent configurations from equilibrated molecular dynamics trajectories and using linear response to calculate the polarizability of each.

\begin{figure}[tbh]
 \includegraphics[height=0.22\textheight,clip]{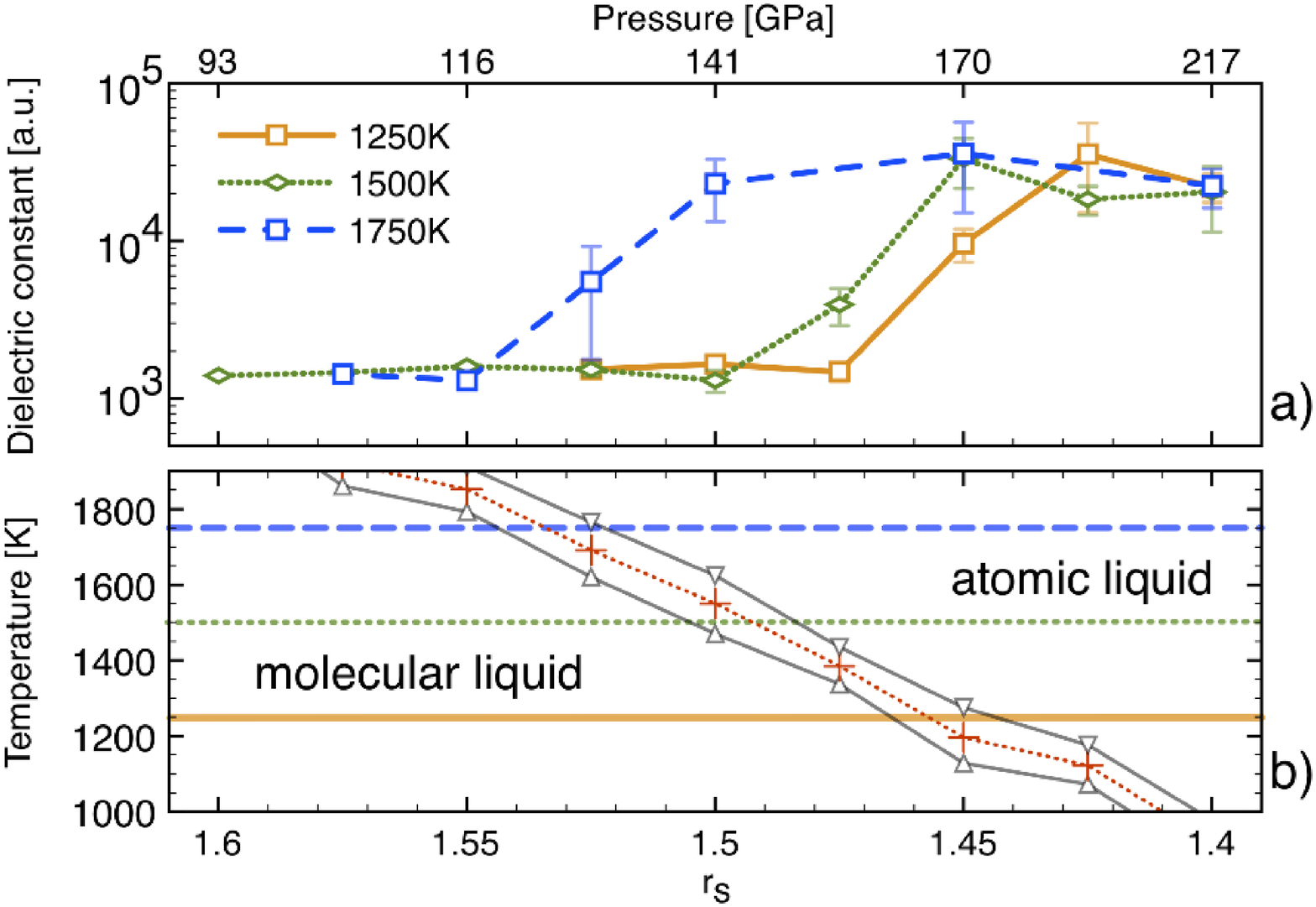} 
  \caption{\label{dielectric} (a) Static dielectric constant of the liquid calculated along compression isotherms. Error bars indicate standard deviation of results from different atomic configurations. The pressure scale corresponds to the 1500 K isotherm. Significant changes in the dielectric properties of the liquid begin at the density corresponding to the onset of dissociation (b). Definition of symbols is the same used in Fig.~\ref{phase}.}
\end{figure}

To map the liquid phase diagram (Fig.~\ref{phase}), we calculate the survival probability, $\Pi(\tau)$, of paired atoms for different $P$, $T$ conditions (over 150 samples, each corresponding to a fully equilibrated molecular dynamics trajectory). A pair is defined to be two H atoms which are mutual nearest neighbors. Given a pair at time $t = 0$, we calculate the probability for the same pair to exist at $t = \tau$ (see Supplimentary Fig.~1). For a purely molecular system, $\Pi(\tau) = 100\%$, while for a purely atomic one $\Pi(\tau) \rightarrow 0\%$ (see Supplimentary Fig.~2). We found that our results are qualitatively insensitive to the parameter $\tau$ (we propose a standard of 10 H$_2$ oscillations). This criterion relates directly to the rate of dissociation in the liquid, and naturally accounts for bond breaking and reforming characteristic of high pressure liquids. The molecular-atomic transition based on $\Pi(\tau) = 50\%$ is shown as a function of $P$ in Fig.~\ref{phase} and as a function of the density parameter $r_s$ (defined by $V/N =\frac{4}{3}\pi(r_{s}a_{0})^{3}$, where $V$ is the volume, $N$ the number of electrons, and $a_o$ is the Bohr radius) in Fig.~\ref{dielectric}(b). Linear extrapolation of the transition line suggests that $r_s \approx 1.25$ is the maximum density where molecules would be stable in a $T = 0$~K liquid (the existence of which has been previously proposed \cite{brovman_spjetp_1972,ashcroft_jpcm_2000,bonev_nature_2004}). The inclusion of quantum zero-point effects is expected to shift the onset of dissociation to lower densities. On the other hand, the stability of hydrogen pairs in a {\em solid} can be enhanced due the presence of intermediate-range order and Friedel oscillations~\cite{nagao_prl_2003}.

To relate the structural changes to the dielectric properties of the liquid, we have computed the static dielectric constant (average of the diagonal elements of the polarizability tensor). As shown in Fig.~\ref{dielectric}, it increases by several orders of magnitude when the system is brought through the dissociation transition. As the probability of molecular survival decreases, the liquid becomes increasingly polarizable. This change is consistent with a transition from an electronically insulating to a conducting state - a metal or a thermally activated semiconductor. 

%
%

The appearance of a maximum in the melting line of hydrogen near 82~GPa has been explained by a softening of intermolecular interactions in the liquid \cite{bonev_nature_2004}, similar to what was observed in the orientationally-ordered phase of the solid (phase III) \cite{moshary_prb_1993}. This has prompted us to investigate the structure of the molecular liquid. We have discovered a new region in the molecular fluid in the vicinity of the melt line turnover that exhibits short range orientational ordering. It is revealed in the spatial distribution function, $SDF(r,\theta)$. Given an H$_2$ molecule, $SDF(r,\theta)$ is defined as the probability for finding a hydrogen atom at a distance $r$ from the molecular center of mass and at an angle $\theta$ with the molecular axis, normalized by the average H number density.

\begin{figure}[tbh]
  \includegraphics[width=0.45\textwidth,clip]{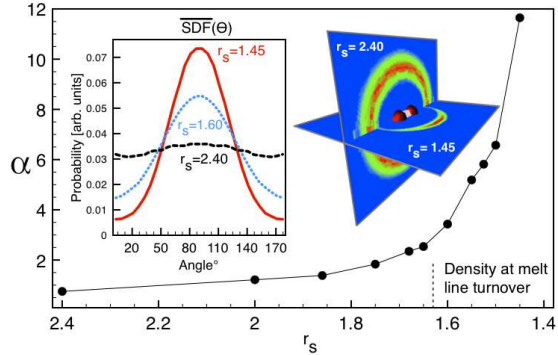}
  \caption{\label{order}Structural order parameter, $\alpha$, and spatial distributions functions, $SDF(r,\theta)$ and $\overline{SDF}(\theta)$ (see text for definitions), along the 1000~K isotherm in molecular H. The density where the melting curve has a maximum is indicated on the plot; it is in the region where $\alpha$ begins to increase rapidly. The graphic shows nearest neighbor contributions to $SDF(r,\theta)$ at low ($r_s= 2.40$, $P=6$~GPa) and high ($r_s = 1.45$, $P=170$~GPa) density. The $\overline{SDF}(\theta)$ further illustrate changes in the angular distribution with density.}
\end{figure}

\begin{figure}[tbh]
  \includegraphics[width=0.45\textwidth,clip]{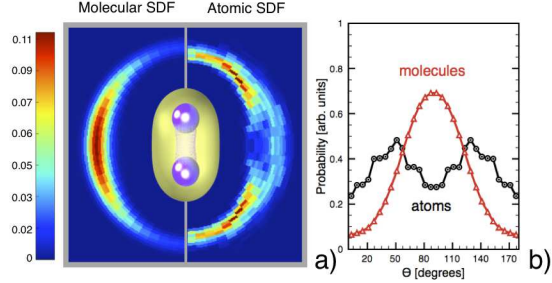}
  \caption{\label{asdf}Molecular and atomic spatial distribution functions, $SDF(r,\theta)$ (a) and $\overline{SDF}(\theta)$ (b); see text for definitions. For clarity, only contributions from nearest neighbor atoms are included. Atoms forming molecules are more likely to reside near the reference molecule's equator. The distribution of unpaired atoms is more even and indicates that they are most likely to be found at $\theta \approx 50^{\circ}$ and $\theta \approx 130^{\circ}$.}
\end{figure}

In Fig.~\ref{order}, we show a graphic of the $SDF$ for two densities along the same 1000~K isotherm; for clarity only the contributions from nearest-neighbors are shown. In the low density case (vertical plane), a molecule's nearest neighbor has an equal probability of being found at any $\theta$. At high pressure (horizontal plane), there is an orientational correlation between neighboring molecules. Particles are less likely to be found at the molecular poles, and more likely to be found near the equators. This can be further seen in the inset of Fig.~\ref{order}, where we have integrated over the radial degree of freedom, $\int \! SDF(r,\theta) \, dr = \overline{SDF}(\theta)$. To describe the evolution of the structural transition under compression, we define an order parameter as $\alpha = \overline{SDF}(\theta=90)/\overline{SDF}(\theta=0)$. A plot of $\alpha$ as a function of $r_s$ along the 1000~K isotherm (Fig.~\ref{asdf}) indicates a rapid increase in the vicinity of the previously reported \cite{bonev_nature_2004, shanti_prl_2008, eremets_jetp_2009} maximum in the melting curve. It is this change in the structure of the liquid, driven by a change in the intermolecular interactions, that is responsible for the turnover in the melting line.

A detailed quantitative description of the emerging orientational order will be presented in a follow up paper. Such analysis is expected to have relevance for identifying the finite (and possible zero) $T$ phases of the solid; similarities with phases 
II and III of H are plausible. Here we mention only that in addition to the described dependence on $\theta$, there is a very strong tendency for neighboring molecules to lie in plane. For $\theta \sim 90^{\circ}$, the angle between molecular axes tends to be about 15$^{\circ}$ (for different $\theta$ it depends slightly on whether the molecules are tilted towards or away from each other). 

One of the most prominent features of dense H is the sharp transition from molecular to atomic liquid. At sufficiently high density, the onset of dissociation leads to an EOS with a negative slope, i.e. $dP/dT |_{V=const} < 0$. There has been considerable debate as to whether the abrupt change in $P$ over a narrow $T$ range is due to a discontinuous ($1^{st}$ order) transition in the liquid and how it is affected by the presence of impurities (e.g. He). In what follows, we show that the anomalous features in the EOS, its density and impurity dependence, and its sensitivity to computational parameters can be understood on the basis of the changes in the liquid structure across the dissociation transition.

In order to describe the liquid structure in the presence of both atoms and molecules, we decompose $SDF(r,\theta)$ into atomic and molecular components. In both cases, $SDF(r,\theta)$ is calculated with respect to reference molecules center of mass. For the atomic $SDF(r,\theta)$, statistics for $r$ and $\theta$ are collected for single atoms only, while for the molecular one, the atoms at $r$ and $\theta$ are paired. In Fig.~\ref{asdf} we show such a decomposition at conditions corresponding to a large degree of local angular structure ($\alpha \sim 10$). At these $P$ and $T$, the liquid has just begun to dissociate ($\Pi(\tau) \approx 99\%$). The molecular $SDF$ is similar to that of a purely molecular liquid, with a peak near the reference molecule's equator. However, the atomic $SDF$ shows that unpaired atoms tend to shift towards the molecular poles. The atomic $\overline{SDF}(\theta)$ is peaked at $\theta \approx 50^{\circ}$ (and $130^{\circ}$) versus $90^{\circ}$ for the molecular $\overline{SDF}(\theta)$ [Fig.~\ref{asdf}(b)]. Similar analysis at low density shows no such angular dependence.

At low $P$, where there is no short-range orientational order, the H$_2$ molecules are freely rotating and their packing is similar to that of spheres. Upon dissociation, the problem becomes that of spheres of different sizes, however, there is no apparent optimization of packing. On the other hand, because of covalent bonding, the effective volume of H$_2$ is less than that of two isolated hydrogen atoms. For these reasons, the dissociation transition at low density is characterized by $dP/dT |_{V=const} > 0$.

At high $P$, the molecules are no longer freely rotating and the packing problem becomes that of prolate spheroids (see graphic in Fig.~\ref{asdf}). As seen Fig.~\ref{asdf}, the molecular orientational order creates voids between them, which can be filled in by single atoms after dissociation. There will be a drop $P$ upon dissociation when the optimization in packing is sufficient to compensate for the higher effective volume of single atoms compared to that of molecules. This picture is consistent with the fact that the sharpness in the EOS increases with density \cite{bonev_prb_2004,miguel,caspersen,vorberger_prb_2007} - our order parameter follows the same trend. Furthermore, it has been noted that diluting the system with neutral impurities such as helium \cite{vorberger_prb_2007} softens the transition. We suggest that performing $SDF$ analysis on high density mixtures of molecular
hydrogen and helium (or other noble gases) will reveal similar angular local order, with the dopant atoms residing within the voids of the liquid. Finally, we note that in a one-dimensional hydrogen liquid, where angular order is not possible, $dP/dT |_{V=const}$ must always be positive.

Packing considerations also provide an explanation for the large sensitivity of the transition to the number of atoms, $N$, in the simulation supercell. We find that the degree of dissociation and related properties depend on $N$ in a non-intuitive way (similar observations have been made by Morales \cite{miguel}). At high densities, $\Pi(\tau)$ oscillates as $N$ is varied from 256, 512, 768, 1024. Surprisingly, simulations conducted with 256 atoms are in better agreement with those done with $N=1024$ than with $N=512$ ($\Pi_{N=128}=0.3338$, $\Pi_{N=256}=0.9842$, $\Pi_{N=512}=0.8328$, $\Pi_{N=768}=0.9660$, and $\Pi_{N=1024}=0.9656$ for $r_s=1.45$, $T=1000 K$). This effect \emph{cannot} be removed with a finer $k$-point sampling of the Brillouin zone. With $N=128$ the error in $P$ is as large as 8\% at $r_s$=1.40, $T$=750 K.

The origin of this initially puzzling behavior is that at a given density, $N$ defines the physical dimensions of the simulation box. Depending on how these dimensions relate to the correlation length of the liquid, different sized cells will have a slight bias towards one phase or another. Ideally one should use a value large enough that this effect is no longer an issue. Given the current cost of using large cells, we report values corresponding to $N=256$, as they are in relatively good agreement with $N=1024$ (the oscillation in $\Pi(\tau)$ is quite damped in this larger cell). Simulations performed with $N=256$ tend to favor the molecular phase relative to that of $N=1024$; this should partially correct for the underestimate of the dissociation barrier due to approximations of the DFT exchange-correlation potential.

The existence of short range angular structure in the liquid, specifically the different arrangement of molecules and atoms, suggests that something similar might also occur in the solid phase. In future searches for solid state structures it would be worthwhile to include variations of a two-component solid, comprised of molecules and atoms. Indeed, work by Pickard and Needs \cite{pickard_natphys_2007} indicates that mixed layered structures, comprised of molecules and atomic graphene-like sheets, remain energetically competitive over a wide range of pressures.

We now turn our discussion to the possiblity of a discontinous (\emph{i.e.} $1^{st}$ order) transition in the liquid. Based on our sampling of the phase diagram, the molecular-atomic transition appears to be continuous up to 200~GPa. We do not find a flat region in the $P(V)$ EOS - a signature of a $1^{st}$ order transition. Our data imply one of three possibilities: the transition is not $1^{st}$ order, it takes place outside of the $P$, $T$ space we considered, or it occurs over a range of densities too small to be resolved by our sampling. We can use our results, however, to establish a connection between the microscopic and macroscopic properties characterizing the transition and provide insight for the physical mechanism that could lead to it being $1^{st}$ order. Let $\delta V$ be the reduction in volume (due to packing) that can be achieved by dissociating a single molecule, while keeping $P$ and $T$ constant (the resulting state need not be in equilibrium). If a $1^{st}$ phase transition exists at these $P$ and $T$, we must have $-P\Delta V = \Delta U - T\Delta S$. Here $\Delta V$, $\Delta U$ and $\Delta S$ are the volume, energy and entropy differences across the transition. If it results in the dissociation of $n_d$ molecules, then we define $E_d \equiv \Delta U/n_d$, which has the physical meaning of dissociation energy. A $1^{st}$ order transition will take place if $n_d \delta V = - \Delta V$, which means $\delta V \approx E_d/P$ (assuming that $T\Delta S/P$ can be neglected, especially at high $P$ and low $T$). We note that $E_d$ \emph{decreases} under compression due to screening, while $\delta V$ \emph{increases} due to more pronounced orientational order. Thus, convergence of these two terms, $\delta V$ and $ E_d/P$, is increasingly likely at high $P$ and low $T$. Conclusive determination of the existence of a critical point will require very fine sampling of the phase diagram below 1000~K and above 200~GPa. 

%
%

We have mapped the molecular-atomic transition of liquid hydrogen over a large pressure range. Our transition line correlates well with changes that can be observed in the electronic properties of the system such as the static dielectric constant. We predict that the liquid demonstrates significant short range orientational ordering. Its development coincides with the turnover in the melt line. The existence of this structural order is responsible for the large dissociation-induced $P$ drop in the EOS. Furthermore, it provides an explanation for the significant finite size effects that are present in this system. A $1^{st}$ order transition in the liquid, if it does exist, will likely occur at $T < $ 1000~K and $P > $ 200~GPa. 

%
%

Work supported by NSERC, CFI, and Killam Trusts. Computational resources provided by ACEnet, Sharcnet, IRM Dalhousie, and Westgrid. We thank E. Schwegler, K. Caspersen, T. Ogitsu, and M. Morales for discussions and communicating unpublished results.


\begin{thebibliography}{23}
\expandafter\ifx\csname natexlab\endcsname\relax\def\natexlab#1{#1}\fi
\expandafter\ifx\csname bibnamefont\endcsname\relax
  \def\bibnamefont#1{#1}\fi
\expandafter\ifx\csname bibfnamefont\endcsname\relax
  \def\bibfnamefont#1{#1}\fi
\expandafter\ifx\csname citenamefont\endcsname\relax
  \def\citenamefont#1{#1}\fi
\expandafter\ifx\csname url\endcsname\relax
  \def\url#1{\texttt{#1}}\fi
\expandafter\ifx\csname urlprefix\endcsname\relax\def\urlprefix{URL }\fi
\providecommand{\bibinfo}[2]{#2}
\providecommand{\eprint}[2][]{\url{#2}}

\bibitem[{\citenamefont{Bonev et~al.}(2004)\citenamefont{Bonev, Schwegler,
  Ogitsu, and Galli}}]{bonev_nature_2004}
\bibinfo{author}{\bibfnamefont{S.~A.} \bibnamefont{Bonev}},
  \bibinfo{author}{\bibfnamefont{E.}~\bibnamefont{Schwegler}},
  \bibinfo{author}{\bibfnamefont{T.}~\bibnamefont{Ogitsu}}, \bibnamefont{and}
  \bibinfo{author}{\bibfnamefont{G.}~\bibnamefont{Galli}},
  \bibinfo{journal}{Nature} \textbf{\bibinfo{volume}{431}},
  \bibinfo{pages}{669} (\bibinfo{year}{2004}).

\bibitem[{\citenamefont{Deemyad and Silvera}(2008)}]{shanti_prl_2008}
\bibinfo{author}{\bibfnamefont{S.}~\bibnamefont{Deemyad}} \bibnamefont{and}
  \bibinfo{author}{\bibfnamefont{I.~F.} \bibnamefont{Silvera}},
  \bibinfo{journal}{Phys. Rev. Lett.} \textbf{\bibinfo{volume}{100}},
  \bibinfo{eid}{155701} (\bibinfo{year}{2008}).

\bibitem[{\citenamefont{Eremets and Trojan}(2009)}]{eremets_jetp_2009}
\bibinfo{author}{\bibfnamefont{M.~I.} \bibnamefont{Eremets}} \bibnamefont{and}
  \bibinfo{author}{\bibfnamefont{I.~A.} \bibnamefont{Trojan}},
  \bibinfo{journal}{JETP Lett.} \textbf{\bibinfo{volume}{89}},
  \bibinfo{pages}{198} (\bibinfo{year}{2009}).

\bibitem[{\citenamefont{{Scandolo}}(2003)}]{scandolo_pnas_2003}
\bibinfo{author}{\bibfnamefont{S.}~\bibnamefont{{Scandolo}}},
  \bibinfo{journal}{PNAS} \textbf{\bibinfo{volume}{100}}, \bibinfo{pages}{3051}
  (\bibinfo{year}{2003}).

\bibitem[{\citenamefont{{Bonev} et~al.}(2004)\citenamefont{{Bonev}, {Militzer},
  and {Galli}}}]{bonev_prb_2004}
\bibinfo{author}{\bibfnamefont{S.~A.} \bibnamefont{{Bonev}}},
  \bibinfo{author}{\bibfnamefont{B.}~\bibnamefont{{Militzer}}},
  \bibnamefont{and} \bibinfo{author}{\bibfnamefont{G.}~\bibnamefont{{Galli}}},
  \bibinfo{journal}{\prb} \textbf{\bibinfo{volume}{69}},
  \bibinfo{pages}{014101} (\bibinfo{year}{2004}).

\bibitem[{\citenamefont{Delaney et~al.}(2006)\citenamefont{Delaney, Pierleoni,
  and Ceperley}}]{delaney_prl_2006}
\bibinfo{author}{\bibfnamefont{K.~T.} \bibnamefont{Delaney}},
  \bibinfo{author}{\bibfnamefont{C.}~\bibnamefont{Pierleoni}},
  \bibnamefont{and} \bibinfo{author}{\bibfnamefont{D.~M.}
  \bibnamefont{Ceperley}}, \bibinfo{journal}{Phys. Rev. Lett.}
  \textbf{\bibinfo{volume}{97}}, \bibinfo{pages}{235702}
  (\bibinfo{year}{2006}).

\bibitem[{\citenamefont{Vorberger{~\it et. al}}(2007)}]{vorberger_prb_2007}
\bibinfo{author}{\bibfnamefont{J.}~\bibnamefont{Vorberger{~\it et. al}}},
  \bibinfo{journal}{Phys. Rev. B} \textbf{\bibinfo{volume}{75}},
  \bibinfo{pages}{024206} (\bibinfo{year}{2007}).

\bibitem[{cas()}]{caspersen}
\bibinfo{note}{K. Caspersen, private communication.}

\bibitem[{\citenamefont{Weir et~al.}(1996)\citenamefont{Weir, Mitchell, and
  Nellis}}]{weir_prl_1996}
\bibinfo{author}{\bibfnamefont{S.~T.} \bibnamefont{Weir}},
  \bibinfo{author}{\bibfnamefont{A.~C.} \bibnamefont{Mitchell}},
  \bibnamefont{and} \bibinfo{author}{\bibfnamefont{W.~J.}
  \bibnamefont{Nellis}}, \bibinfo{journal}{Phys. Rev. Lett.}
  \textbf{\bibinfo{volume}{76}}, \bibinfo{pages}{1860} (\bibinfo{year}{1996}).

\bibitem[{\citenamefont{Nellis et~al.}(1999)\citenamefont{Nellis, Weir, and
  Mitchell}}]{nellis_prb_1999}
\bibinfo{author}{\bibfnamefont{W.~J.} \bibnamefont{Nellis}},
  \bibinfo{author}{\bibfnamefont{S.~T.} \bibnamefont{Weir}}, \bibnamefont{and}
  \bibinfo{author}{\bibfnamefont{A.~C.} \bibnamefont{Mitchell}},
  \bibinfo{journal}{Phys. Rev. B} \textbf{\bibinfo{volume}{59}},
  \bibinfo{pages}{3434} (\bibinfo{year}{1999}).

\bibitem[{\citenamefont{Cassen{~\it et. al}}(2006)}]{cassen_springer_2006}
\bibinfo{author}{\bibfnamefont{P.}~\bibnamefont{Cassen{~\it et. al}}},
  \emph{\bibinfo{title}{Extrasolar planets}} (\bibinfo{publisher}{Springer},
  \bibinfo{year}{2006}).

\bibitem[{mig()}]{miguel}
\bibinfo{note}{M. Morales, private communication.}

\bibitem[{\citenamefont{Datchi et~al.}(2000)\citenamefont{Datchi, Loubeyre, and
  LeToullec}}]{datchi_prb_2000}
\bibinfo{author}{\bibfnamefont{F.}~\bibnamefont{Datchi}},
  \bibinfo{author}{\bibfnamefont{P.}~\bibnamefont{Loubeyre}}, \bibnamefont{and}
  \bibinfo{author}{\bibfnamefont{R.}~\bibnamefont{LeToullec}},
  \bibinfo{journal}{Phys. Rev. B} \textbf{\bibinfo{volume}{61}},
  \bibinfo{pages}{6535} (\bibinfo{year}{2000}).

\bibitem[{\citenamefont{Gregoryanz{~\it et. al}}(2003)}]{gregoryanz_prl_2003}
\bibinfo{author}{\bibfnamefont{E.}~\bibnamefont{Gregoryanz{~\it et. al}}},
  \bibinfo{journal}{Phys. Rev. Lett.} \textbf{\bibinfo{volume}{90}},
  \bibinfo{pages}{175701} (\bibinfo{year}{2003}).

\bibitem[{\citenamefont{Attaccalite and Sorella}(2008)}]{attaccalite_prl_2008}
\bibinfo{author}{\bibfnamefont{C.}~\bibnamefont{Attaccalite}} \bibnamefont{and}
  \bibinfo{author}{\bibfnamefont{S.}~\bibnamefont{Sorella}},
  \bibinfo{journal}{Phys. Rev. Lett} \textbf{\bibinfo{volume}{100}},
  \bibinfo{pages}{114501} (\bibinfo{year}{2008}).

\bibitem[{\citenamefont{Kohn and Sham}(1965)}]{kohn_sham}
\bibinfo{author}{\bibfnamefont{W.}~\bibnamefont{Kohn}} \bibnamefont{and}
  \bibinfo{author}{\bibfnamefont{L.~J.} \bibnamefont{Sham}},
  \bibinfo{journal}{Phys. Rev.} \textbf{\bibinfo{volume}{140}},
  \bibinfo{pages}{A1133} (\bibinfo{year}{1965}).

\bibitem[{\citenamefont{Perdew et~al.}(1996)\citenamefont{Perdew, Burke, and
  Ernzerhof}}]{PBE}
\bibinfo{author}{\bibfnamefont{J.~P.} \bibnamefont{Perdew}},
  \bibinfo{author}{\bibfnamefont{K.}~\bibnamefont{Burke}}, \bibnamefont{and}
  \bibinfo{author}{\bibfnamefont{M.}~\bibnamefont{Ernzerhof}},
  \bibinfo{journal}{Phys. Rev. Lett.} \textbf{\bibinfo{volume}{77}},
  \bibinfo{pages}{3865} (\bibinfo{year}{1996}).

\bibitem[{cpm()}]{cpmd}
\bibinfo{note}{CPMD v 3.11.1, http://www.cpmd.org.}

\bibitem[{\citenamefont{Brovman et~al.}(1972)\citenamefont{Brovman, Kagan, and
  Kholas}}]{brovman_spjetp_1972}
\bibinfo{author}{\bibfnamefont{E.~G.} \bibnamefont{Brovman}},
  \bibinfo{author}{\bibfnamefont{Y.}~\bibnamefont{Kagan}}, \bibnamefont{and}
  \bibinfo{author}{\bibfnamefont{A.}~\bibnamefont{Kholas}},
  \bibinfo{journal}{Sov. Phys. JETP} \textbf{\bibinfo{volume}{35}},
  \bibinfo{pages}{783} (\bibinfo{year}{1972}).

\bibitem[{\citenamefont{Ashcroft}(2000)}]{ashcroft_jpcm_2000}
\bibinfo{author}{\bibfnamefont{N.~W.} \bibnamefont{Ashcroft}},
  \bibinfo{journal}{J. Phys.: Cond. Matt.} \textbf{\bibinfo{volume}{12}},
  \bibinfo{pages}{A129} (\bibinfo{year}{2000}).

\bibitem[{\citenamefont{Nagao{~\it et. al}}(2003)}]{nagao_prl_2003}
\bibinfo{author}{\bibfnamefont{K.}~\bibnamefont{Nagao{~\it et. al}}},
  \bibinfo{journal}{Phys. Rev. Lett.} \textbf{\bibinfo{volume}{90}},
  \bibinfo{pages}{035501} (\bibinfo{year}{2003}).

\bibitem[{\citenamefont{Moshary et~al.}(1993)\citenamefont{Moshary, Chen, and
  Silvera}}]{moshary_prb_1993}
\bibinfo{author}{\bibfnamefont{F.}~\bibnamefont{Moshary}},
  \bibinfo{author}{\bibfnamefont{N.~H.} \bibnamefont{Chen}}, \bibnamefont{and}
  \bibinfo{author}{\bibfnamefont{I.~F.} \bibnamefont{Silvera}},
  \bibinfo{journal}{Phys. Rev. B} \textbf{\bibinfo{volume}{48}},
  \bibinfo{pages}{12613} (\bibinfo{year}{1993}).

\bibitem[{\citenamefont{Pickard and Needs}(2007)}]{pickard_natphys_2007}
\bibinfo{author}{\bibfnamefont{C.~J.} \bibnamefont{Pickard}} \bibnamefont{and}
  \bibinfo{author}{\bibfnamefont{R.~J.} \bibnamefont{Needs}},
  \bibinfo{journal}{Nat. Phys.} \textbf{\bibinfo{volume}{3}},
  \bibinfo{pages}{473} (\bibinfo{year}{2007}).

\end{thebibliography}

\end{document}